# MANDATORY DISCLOSURE OF STANDARDIZED SUSTAINABILITY METRICS: THE CASE OF THE EU TAXONOMY REGULATION[*]

Marvin Nipper[†]   Andreas Ostermaier[‡]   Jochen Theis[§]

**Abstract**

Sustainability reporting enables investors to make informed decisions and is hoped to facilitate the transition to a green economy. The European Union's taxonomy regulation enacts rules to discern sustainable activities and determine the resulting green revenue, whose disclosure is mandatory for many companies. In an experiment, we explore how this standardized metric is received by investors relative to a sustainability rating. We find that green revenue affects the investment probability more than the rating if the two metrics disagree. If they agree, a strong rating has an incremental effect on the investment probability. The effects are robust to variation in investors' attitudes. Our findings imply that a mandatory standardized sustainability metric is an effective means of channeling investment, which complements rather than substitutes sustainability ratings.

---

[*] We thank Hans Frimor, Răzvan Ghiță, Qiang Guo, Seung Lee, Rainer Lueg, Peter Schäfer, Matthias Schmidt, and Dennis van Liempd for helpful comments. We gratefully acknowledge generous funding from the Fynske Købstæders Fond.
[†] University of Duisburg-Essen, Forsthausweg 2, 47057 Duisburg, Germany, marvin.nipper@uni-due.de.
[‡] University of Southern Denmark, Campusvej 55, 5230 Odense, Denmark, ostermaier@sam.sdu.dk.
[§] University of Southern Denmark, Universitetsparken 1, 6000 Kolding, Denmark, jot@sam.sdu.dk.

# MANDATORY DISCLOSURE OF STANDARDIZED SUSTAINABILITY METRICS: THE CASE OF THE EU TAXONOMY REGULATION

## I. INTRODUCTION

Sustainability reporting helps investors make informed investment decisions. It can channel funding to companies that pursue sustainable business and therefore facilitate the transition to a green economy. The European Union has enacted a taxonomy regulation (Regulation 2020/852) to standardize sustainability reporting and reduce greenwashing. Many companies are henceforth required to disclose the portions of their revenue, capital expenditures, and operating expenditures that are associated with sustainable business, as defined by the taxonomy rules. The government provision of standardized sustainability metrics, similar to financial metrics, sets a precedent for standard setting and regulation. However, green revenue and expenditures compete with third-party assured sustainability ratings. A company may thus present multiple, incongruent metrics, which all pretend to measure its sustainability. It is yet unclear how investors will receive green revenue, relative to sustainability ratings. We utilize an experiment to examine how the disclosure of green revenue and sustainability ratings combine to influence investors' decisions. Our findings suggest that a mandatory standardized metric like that of the European Union effectively channels investment, complementing rather than substituting sustainability ratings.

The disclosure of green revenue is mandatory for companies that fall under the European Union's non-financial reporting directive (Directive 2014/95/EU), and the European Commission (2021) has proposed a corporate sustainability reporting directive that will extend the scope of companies to which the regulation applies. Sustainability ratings are, in turn, voluntary. Hence, a company that fails or knows it will fail to obtain a decent rating can decide to not report any rating, but it must report its green revenue, even if it is zero. We first focus on the cases where a company discloses any combination of either much or little green revenue along with either a favorable or no sustainability rating. An environmentally sustainable business promises higher financial performance for reasons such as product and labor market benefits as well as reduced litigation risk and compliance cost, and it is therefore more interesting for investors. In addition, some investors are willing to pay a premium for stock in sustainable business (Richardson and Welker 2001, Dhaliwal et al. 2011, Matsumura et al. 2014, Plumlee et al. 2015). Taken together, these arguments lead us to hypothesize that investors are more likely to invest in a company that discloses much rather than little green revenue or a high rather than no sustainability rating.

Moving beyond the individual effects of green revenue and sustainability ratings, we examine how these combine to influence investors. The taxonomy seeks to reduce greenwashing, which sustainability ratings are susceptible to (Scalet and Kelly 2010, Windolph 2011). Sustainability ratings are also heterogeneous (Christensen et al. 2022), and thus companies can cherry-pick favorable ratings. Companies that have little green revenue to disclose may attempt to combine this bad news with a favorable sustainability rating. We predict that a good sustainability rating mitigates but does not make up for a lack of green revenue. Conversely, a company whose revenue is green can easily obtain a high sustainability rating in addition. While the rating further underscores the company's sustainability, it does not convey new information in this case. Hence, the question arises of whether investors will discount the rating. Finally, we consider the case where the disclosure of green revenue is voluntary. Companies that are outside the scope of the taxonomy regulation are not required to but can disclose their green revenue. We hypothesize that the voluntary disclosure of green revenue is worthwhile if it is high. Conversely, we expect that it is worse for a company to disclose little green revenue than to not disclose its green revenue.

To test our predictions, we conducted an online experiment on CloudResearch, where our participants took the role of private investors and indicated their likelihood to invest in a company.

We manipulated, within subjects, the information about the company's green revenue and sustainability rating, each at three levels. Thus, the company reported either much or little green revenue, or it did not disclose its green revenue. Likewise, the company reported either a high or a low sustainability rating, or it did not report any rating. In addition, we manipulated whether the company reported a high or low financial performance. Thus, each investor indicated her or his investment probability in eighteen (= 3 × 3 × 2) cases. The instructions explained the concepts of green revenue and the taxonomy as well as sustainability ratings without any reference to the European Union or any specific rating agency. An attention check ensured that our participants understood this information and could hence reasonably indicate their investment probability in each case. After they had made their investment decisions, our investors answered questions about their attitude to the protection of the environment and to government regulation of businesses and markets. The survey concluded with questions to collect minimal demographic data.

    The results of our experiment support our predictions. First, we find that the disclosure of much green revenue and of a high sustainability rating, against the baselines of the disclosure of little green revenue and no disclosure of any sustainability rating, each raise the probability of investment. Second, a high sustainability rating mitigates the negative effect of disclosing little green revenue on the investment probability but does not level it out. The answer to our question is that a high sustainability rating has the same effect on the investment probability whether it combines with much or little green revenue. That is, investors count rather than discount favorable bits of information. Third, the (voluntary) disclosure of much green revenue raises the investment probability relative to non-disclosure, and the disclosure of little green revenue diminishes it. Additional analyses reveal interactions with pro-environment and pro-government attitudes. A strong pro-environment attitude reinforces the effect of the disclosure of much green revenue or a

high rating. A strong pro-government attitude reinforces the former but not the latter. Moreover, it increases the investor's reliance on green revenue relative to a sustainability rating. The observed effects are more pronounced if financial performance is high than if it is low.

Governments and public authorities exert increasing regulatory pressure on businesses to mitigate climate change. Our study contributes in at least two ways to our understanding of how sustainability reporting affects business decisions. First, the disclosure of green revenue increases the investment probability, which suggests that regulation can channel investments to sustainable business. Indeed, investors respond to the disclosure of green revenue more than to a high sustainability rating. For this differential effect, it is important to keep in mind that the regulation makes the disclosure of green revenue mandatory, even if the company has none to report. Thus, the baseline green revenue is zero by default, whereas companies can decide to not disclose a poor sustainability rating. Moreover, the moderating effect of the pro-government attitude also supports the argument that investors appreciate the governmental backing of the taxonomy rules. While much of the literature is about voluntary disclosure, recent research documents real effects of sustainability reporting mandates in the European Union and the United Kingdom, which argues for regulation (Downar et al. 2021, Fiechter et al. 2022). Contrasting a regulatory and a non-regulatory approach to sustainability reporting, our study also makes a case for regulation.

From a managerial viewpoint, our study holds two lessons for companies that fall under the taxonomy or a similar regulation. First, any company that wants to improve its sustainability needs to decide whether to change its operations to avoid making any impact or to offset the impact that it will make. Operational change turns the company's revenue green and normally results in a high sustainability rating as a result. While it thus seems sufficient to disclose green revenue, it still pays off to obtain a good sustainability rating in addition. Investors do not fully appreciate the dependence of the sustainability rating on green revenue and are willing to additionally reward the high rating. Second, unlike ratings, the taxonomy does not admit offsetting, which may therefore increase the company's rating but not its green revenue. A good sustainability rating, which can be obtained by offsetting, can mitigate the damage that results from the mandatory disclosure of the little green revenue that the company has, but it does not level it out. While offsetting may still be less costly for a company than operational change, the mandatory disclosure of green revenue increases this cost. Incidentally, if disclosure is voluntary for a company, the disclosure of much green revenue is better and the disclosure of little green revenue worse than non-disclosure.

The remainder of this paper unfolds in four steps. We first provide more information about the taxonomy regulation and develop our hypotheses (Section II). We then explain our experimental design (Section III) and present our results (Section IV). Finally, we conclude with a summary and discussion of our findings (Section V).

## II. BACKGROUND AND HYPOTHESES

**Background**

In response to climate change, the European Union initiated a comprehensive set of policies to make the European economy climate-neutral (European Commission 2018). Sustainability reporting is a centerpiece of this so-called European Green Deal. The taxonomy regulation (Regulation 2020/852), along with its delegated regulations and technical annexes, establishes criteria to for economic activities to qualify as environmentally sustainable. The purpose of the regulation is to channel investments to these activities. Having identified their sustainable activities, companies disclose the revenue, capital expenditures, and operating expenditures related to these. The taxonomy regulation implements rules to standardize sustainability reporting and, with the taxonomy-aligned or "green" revenue and expenditures, defines metrics to quantify the company's performance, which enables investors to make informed decisions. In particular, the climate delegated act (Delegated Regulation 2021/2139) requires companies to disclose revenue and expenditures associated with economic activities that help mitigate climate change or that help society to adapt to climate change. This includes, for example, activities that stabilize the level of greenhouse gas concentration or that shield people against the adverse impact of greenhouse gas.

Starting with the non-financial reporting directive (Directive 2014/95/EU), the European Union has enforced sustainability reporting requirements which have produced sizable real effects (Fiechter et al. 2022). The taxonomy regulation goes beyond this directive, which required large companies to report on their sustainability but gave them much discretion on what and how to report. For example, companies that fell under the directive were free to decide whether to follow the European Commission's (2017) non-binding guidelines or some other set of sustainability reporting standards of their choosing (e.g., the GRI framework). The requirements and metrics established by the taxonomy regulation differ from extant standards. First, the taxonomy regulation defines a small set of metrics to measure a company's sustainability, such as green revenue, and makes their disclosure mandatory. Second, it specifies detailed rules for how to compute these metrics, which are transparently stipulated by delegated regulations. The taxonomy rules resemble government-backed accounting standards. Beyond the particular case of a European regulation, they offer lessons for the design of reporting standards, such as the SEC's (2022) proposed rules to enhance and standardize climate-related disclosures for investors.[1]

The non-financial reporting directive (Directive 2014/95/EU) has required large European companies to report on their sustainability since 2017. In the U.S., environmental regulations have, indirectly, entailed climate-related disclosure requirements (SEC 2010). However, neither direct nor indirect prior regulation has obliged companies to summarize the sustainability of their business in metrics like green revenue and expenditures. Sustainability ratings by independent rating agencies have long been a way for companies to voluntarily document their sustainability. There are multiple rating agencies that sometimes offer multiple different ratings, which diverge remarkably (Christensen et al. 2022). The plethora of incongruent ratings allows companies to cherry-pick and choose which ratings to solicit and disclose. Incidentally, companies can always not solicit or disclose any rating. Unsurprisingly, it is uncommon to see a company report a poor sustainability rating. By the taxonomy regulation, in turn, revenue and expenditures are "brown" by default. If a company fails to show, for example, that some portion of its revenue is green, it will have to report zero green revenue. Thus, the mandatory disclosure of green revenue and expenditures is markedly different from the voluntary disclosure of sustainability ratings.

---

[1] For example, the SEC (2022) considers requiring companies to disclose "the amount of capital expenditure deployed toward climate-related risks and opportunities" (p. 54), which resembles green expenditures according to the European taxonomy regulation.

There are different ways for companies to manage their environmental impact. On the one hand, they can try to make their unsustainable operations more sustainable or abandon them to reduce their own impact. On the other hand, they can continue such operations but compensate their impact by purchasing offset credit from third parties (Johnson et al. 2020). Thus, a company keeps emitting greenhouse gas but helps eliminate emissions elsewhere. Operational change, in turn, can refer to core or peripheral business activities (Guiral et al. 2020). The purpose of the taxonomy regulation is operational change. The taxonomy is a comprehensive list of activities that the European Union considers sustainable. Having analyzed their own activities, companies determine whether these align with any of those in taxonomy. In case of a match, the revenue from and expenditures for this activity count as green. Sustainability ratings allow offsetting, such as the purchase of carbon offset credits, which is widely used by companies to attain sustainability targets (Dhanda and Hartman 2011; Johnson et al. 2020). By contrast, the taxonomy rules out offsetting and is hoped to close potential loopholes for greenwashing.[2] The taxonomy's rule-based approach makes green revenue and expenditures tamper-proof performance measures.[3]

**Individual effects of green revenue and sustainability rating disclosure**

Sustainability like financial reports contain decision-relevant information for investors. The literature proposes several arguments for why the disclosure of information about a company's sustainability influences investors. These fall into two broad categories. First, the disclosure of good performance leads investors to expect (positive) cashflow effects to arise from the company's responsible behavior (Richardson et a. 1999). Specifically, such behavior reduces the risk of litigation; it helps prevent government regulation and the related cost of compliance; and it appeals to certain consumers and consumers and therefore carries product and factor market benefits—or, conversely, it helps avoid negative publicity and boycotts (Richardson et al. 1999, Dhaliwal et al. 2011, Matsumura et al. 2014). Taken together, these arguments imply that responsible corporate behavior, through lower costs or higher revenues, promises higher financial performance and thus makes the investment more worthwhile (Richardson and Welker 2001, Al-Tuwaijri et al. 2004). Conversely, if a company chooses to not disclose information about its sustainability, investors

---

[2] The influential group of experts that assists the European Commission in developing the taxonomy-related regulations clarifies that offsetting cannot align an activity with the taxonomy. Both activities are distinct, and the impacting and offsetting activities must not be "netted out" (see Platform on Sustainable Finance, 2021, pp. 60–65, for a detailed discussion of the matter).

The taxonomy recognizes activities that do not substantially contribute to climate change mitigation and adaptation but enable activities that do. An example is the manufacture of batteries, which are needed to run electric vehicles (Delegated Regulation 2021/2139, Annex I, Section 3.4). Again, such activities must be listed in the taxonomy to count as green.

[3] There are plenty of examples of socially irresponsible companies being included in sustainable stock indices, such as the Dow Jones Sustainability Index (Arribas et al. 2019). The commercial nature of sustainability ratings also opens the door to manipulation and insufficient verification of information (Scalet and Kelly 2010, Windolph 2011).

will not only make inferences about its sustainability; they may also get suspicious and take non-disclosure as a bad signal, and they may penalize the company (Matsamura et al. 2014).

Second, some investors are willing to pay a premium for stock from socially responsible corporations, regardless of whether that responsible behavior pays off. Indeed, many investors have prosocial and especially pro-environment attitudes, and the emergence of ethical, green, and other socially responsible investing shows that such investors "vote" with their dollars (Richardson and Welker 2001). By this argument, companies that are sustainable or otherwise socially responsible can attract a larger pool of solvent investors (Dhaliwal et al. 2011, El Ghoul et al. 2011). The effect of investors' non-financial preferences is arguably reinforced by biases. Thus, investors have been found to respond affectively to both good and bad social performance and consequently to tend to over-evaluate or under-evaluate the fundamental value of companies (Elliott et al. 2014, Guiral et al. 2020). Investors who do not share such social preferences still need to account for them in their investment decisions, as the second category of cashflow arguments implies. In summary, the two categories of arguments suggest that corporations that disclose good sustainability performance are more appealing to investors and thus have a lower cost of equity capital (Richardson et al. 1999, El Ghoul et al. 2011, Plumlee et al. 2015).

The evidence on whether voluntary social disclosure reduces the cost of equity is nonetheless mixed. Thus, Richardson and Welker (2001) find a positive association between social disclosure and the cost of equity; Clarkson et al. (2013) do not find any association; Dhaliwal et al. (2011) find the predicted negative association. Recent studies rather argue for a negative association, too (Cheng et al. 2014, Plumlee et al. 2015, Egginton and McBrayer 2018). Measurement issues are a potential reason for the mixed evidence. Indeed, Plumlee et al. (2015) show that the association between disclosure and the cost of equity depends on the valence of the information disclosed. Positive disclosure is negatively associated with the cost of equity, negative disclosure is positively associated with it. Matsumura et al. (2014) find a negative association between the volume of carbon emissions disclosed and firm value, although disclosure is voluntary and the companies with the highest emission volumes do not arguably disclose these. Still, the residual variation is enough to establish the association. Likewise, Holm and Rikhardson (2008) find, in an experiment, that investors allocate more funding to a company that discloses positive environmental information than a company that does not disclose such information.

A sustainability rating provides third-party assured summary information about a company's sustainability. Non-financial reports are mandatory in some jurisdictions and voluntary in others. However, even the mandatory non-financial reports in the European Union do not need to be assured (yet).[4] Being assured by an independent rating agency, a sustainability rating appears more credible than a non-financial report without assurance, and it is voluntary even when non-financial reporting is mandatory (Simnett et al. 2009, Reimsbach et al. 2018). Based on the above arguments along with the available empirical evidence, we predict that a high sustainability rating increases the probability of investment. Since it is voluntary for companies to report a sustainability rating, the baseline to measure the effect of a high rating is no rating. The taxonomy-aligned green revenue also summarizes a company's sustainability. Although not assured (yet), its measurement follows strict rules, and companies can be held liable for mismeasurement. Hence, we predict that much green revenue, just like a high sustainability rating, increases the probability of investment. However, since the disclosure of green revenue is mandatory, we benchmark the effect of disclosing much green revenue against the disclosure of little green revenue.

---

[4] The European Commission's (2021) recent proposal for a corporate sustainability reporting directive mandates limited assurance of sustainability reports.

**H1:** Investors are more likely to invest in companies that report much green revenue than in companies that report little green revenue.

**H2:** Investors are more likely to invest in companies that report a high sustainability rating than in companies that do not report a sustainability rating.

**Combined effect of green revenue and sustainability rating disclosure**

The effect of reporting a sustainability rating together with green revenue on the investment probability is interesting from both a managerial and a regulatory viewpoint. As noted earlier, there are two ways for companies to manage their environmental impact: operational change and offsetting (Johnson et al. 2020). While individual companies can abandon unsustainable operations that cannot be changed, this is infeasible for society as a whole. This is a case for offsetting (Johnson et al. 2020). Likewise, in addition to activities that contribute substantially to climate mitigation and adaptation, the taxonomy regulation also allows enabling and transitional activities, such as the manufacture of batteries (Delegated Regulation 2021/2139, Annex I, Section 3.4) or the manufacture of aluminum, iron, and steel (Sections 3.8–3.9). Nonetheless, the regulation does favor operational change, and it does not allow offsetting to turn unsustainable into sustainable activities. Although operational change is also preferred by investors (Johnson et al. 2020), offsetting is very popular with companies. For example, the market for carbon offsetting has been steadily growing for decades (Donofrio et al. 2021). Hence, the taxonomy regulation prohibits many companies from using their preferred means of increasing their sustainability.

Since the disclosure of taxonomy-aligned activities is also mandatory, companies that fall under the regulation have no choice but to report their green revenue, no matter how little they have to report. Given the negative effect on investment probability that the disclosure of little green revenue entails according to H1, the taxonomy regulation builds up much pressure for operational change. However, while green revenue is hard to temper with because of the rule-based approach of the taxonomy, companies can at least supplement the disclosure of little green revenue with a high sustainability rating. Unlike green revenue, sustainability ratings factor in offsetting efforts. Moreover, the plethora of ratings gives companies a chance to find ratings that attests sustainability (Christensen et al. 2022). In line with H2, we expect that a good sustainability rating increases the investment probability and thus mitigates the adverse effect of the simultaneous disclosure of little green revenue. That said, we do expect that a good sustainability rating does not neutralize this effect. The combination of a high rating and little green revenue is contradictory information, which unsettles investors and discourages them from investing relative to the benchmark case, where the company discloses that much of its revenue is green but no sustainability rating.[5]

> **H3:** The positive effect of reporting a high sustainability rating does not fully compensate the negative effect of reporting little green revenue on the investment probability.

The disclosure of a sustainability rating is an option not only for companies that have little green revenue to disclose and want to mitigate the adverse effect of this mandatory unfavorable disclosure on the investment probability. On the contrary, a company that has much green revenue may also solicit and report a sustainability rating. Returning to the two ways of managing environmental impact (Johnson et al. 2020), the taxonomy favors operational change over

---

[5] The irrelevance of offsetting for green revenue is a likely reason for the combination of little green revenue and a high sustainability rating to occur from the company's viewpoint. Turning to the investors' viewpoint, our argument for H3 does not assume that investors understand the distinction between operational change and offsetting.

There are examples of reports that disclose taxonomy-aligned information but challenge the regulation. For example, Continental (2021) criticizes its "ambiguous wording" (p. 44). Of course, recent reports also reflect discomfort about the hasty enforcement. It is too early to tell whether companies are going to continue presenting similar arguments.

offsetting, whereas sustainability ratings recognize both. A company with much green revenue can thus normally obtain a high sustainability rating without any extra cost for offsetting. On the one hand, it is straightforward to communicate this additional signal and thereby stress the company's commitment to sustainability. On the other hand, the sustainability rating provides indeed redundant information in this case. Hence, investors might discount the additional information, particularly those who understand that green revenue and the sustainability rating are both driven by the same sustainable operations. Consequently, a high sustainability rating should have less of an incremental effect on the investment probability if combined with the disclosure of much green revenue than if combined with little green revenue, in which case it is arguably informative.

That said, investors are susceptible to biases. In particular, they are tempted to treat green revenue and sustainability ratings as independent although they are not. As a result, they would overweight a high rating in their decision on whether to invest in a company with much green revenue. Koonce and Lipe (2017) propose a counting heuristic to explain the effect of earnings surprises on investors' valuation of firms. The count of positive versus negative surprises strongly influences valuation and often dominates the magnitudes of the surprises. The same heuristic explains the effect of guidance frequency on investor judgments (Tang and Venktaraman 2018). Likewise, investors have been found to apprehend correlations heuristically by counting comovements between stock returns. Again, the count dominates the proper correlation, and there is more evidence in the finance literature that investors struggle to understand the concept of correlation (Ungeheuer and Weber 2021). Summing up, the counting heuristic suggests that investors neglect the correlation between green revenue and the sustainability rating and argues for the same incremental effect of a high rating whether much or little revenue is green. Taken all together, the arguments rule out a positive interaction and suggest a weak directional prediction.

**H4:** The incremental effect of reporting a high sustainability rating on the investment probability is weaker if a company reports much green revenue than if it reports little green revenue, or it is the same in both cases.

**Effect of voluntary versus mandatory disclosure of green revenue**

H1 benchmarks the disclosure of much green revenue against that of little because the taxonomy regulation makes disclosure mandatory for many companies. However, although the European Commission (2021) proposes to extend the scope of the regulation, the disclosure of green revenue and expenditures remains voluntary for smaller and non-European companies. Thus, the case of non-disclosure is worth considering from two viewpoints. First, if disclosure increases the investment probability compared to non-disclosure, companies that do not fall under the regulation have an incentive to voluntarily disclose their green revenue. From the managerial viewpoint, it is therefore interesting to compare the investment probability in the cases where the company discloses much green revenue and where it does not disclose its green revenue. Second, from a regulatory viewpoint, mandatory disclosure is hoped to create an incentive for operational change. The prospect of being penalized for the mandatory disclosure of little green revenue reinforces this incentive. This argument implies that companies would not voluntarily disclose their green revenue if they have little, and it calls for the (hypothetical) comparison of the cases where the company discloses little green revenue and where it does not disclose its green revenue.

For H2, we argued that the voluntary disclosure of a high sustainability rating increases the probability of investment relative to non-disclosure. The same argument applies to the voluntary disclosure of green revenue, which leads us to hypothesize that the disclosure of much green revenue also increases the investment probability relative to non-disclosure. Turning to the second comparison, there are, on the one hand, arguments to suggest that unfavorable disclosure is no worse or possibly even better than non-disclosure. Most importantly, non-disclosure is an adverse signal for investors. If a company does not disclose its green revenue, investors may default to the assumption that it has little or no green revenue to report; consequently, they would be indifferent between unfavorable disclosure and non-disclosure. Some of the earlier research on (voluntary) social disclosure disregards the content of the disclosure (Richardson and Welker 2001, Dhaliwal et al. 2011, Clarkson et al. 2013). Matsamura et al. (2014) even find that the median value of firms which do not disclose their carbon emissions is lower than that of firms which disclose them. Although the volume of carbon emissions is negatively associated with firm value, it thus seems that non-disclosure can be worse than disclosure (Matsamura et al. 2014).

On the other hand, the arguments presented for H1 and H2 imply that the disclosure of unfavorable information is worse than non-disclosure. For example, if companies should want to showcase their socially responsible behavior to appease regulators and argue for their reduced risk of litigation, the disclosure of irresponsible behavior calls in turn for regulation, invites litigation, supports competitors' green marketing, and facilitates activist campaigns (Matsamura et al. 2014). Likewise, if favorable social disclosure allows companies to capitalize on investors' prosocial attitudes and decrease their cost of equity, then unfavorable disclosure makes the pool of investors shrink and increases the cost of equity. In Matsamura et al. (2014), the volume of carbon emissions is still negatively associated with firm value among disclosers, and we cannot tell whether non-disclosers' low firm value would be even lower if they disclosed their arguably high emission volumes. Similarly, Plumlee et al. (2015) find a differential effect of good versus bad news on the cost of equity, and "sin" industries, such as the tobacco and nuclear industries, have a higher cost of equity (El Ghoul et al. 2011). Hence, we still expect and hypothesize that the disclosure of little green revenue decreases the investment probability compared to non-disclosure.

**H5:** Investors are more likely to invest in companies that report much green revenue than in companies that do not report their green revenue.

**H6:** Investors are less likely to invest in companies that report little green revenue than in companies that do not report green revenue.

## III. EXPERIMENT

**Participants**

We recruited 180 participants from Amazon.com's Cloud Research platform (formerly known as Amazon Mechanical Turk) to proxy for private investors. We required that our participants reside in the U.S., have completed at least 100 other Cloud Research or Mechanical Turk assignments, and have at least a 90-percent approval rate from prior assignments. Furthermore, we confined the pool of participants to those approved by Cloud Research to ensure a high quality of our sample. 99 participants passed the attention check (see below). 59 of these 99 participants were male; 39, female. One participant did not state her or his gender. The participants' age averaged 40.5 years (with a standard deviation of 11.0). About 63 percent had a Bachelor's or a higher academic degree. Our participants earned $3.50 and took on average less than twelve minutes (with a standard deviation of nine) to complete our survey. This corresponds to an hourly wage of about $17.50, which is well above the minimum wage in the U.S.

While our research is inspired by the European Union's taxonomy regulation, we are interested in the effect of design choices that this regulation instantiates (i.e., the mandatory disclosure of a sustainability metric that results from rule-based standards). These design choices are not inherently linked to the taxonomy regulation and are of more general relevance. Therefore, we phrased the experimental instructions neutrally and never referred to the European Union. We administered our experiment to U.S. residents because the usage of Cloud Research is more prevalent in the U.S. Incidentally, Cloud Research participants have been found to be representative of the U.S. population (Berinsky et al. 2012; Buhrmester et al. 2011; Paolacci et al. 2010). They perform similarly to in-laboratory participants in intelligence tests (Buchheit et al. 2017) and problem-solving and learning tasks (Crump et al. 2013). Consequently, Cloud Research experiments produce similar results as laboratory experiments (Casler et al. 2013; Horton et al. 2011). We obtained ethical approval for our study prior to conducting it.

**Task and procedure**

We employ a within-subjects design where the participants take the role of a private investor in search of new investment opportunities. The investor indicates her or his probability to invest in a company. We manipulate the information about that company, resulting in eighteen cases, which are presented to the investor one by one. While we follow a similar procedure as in a conjoint experiment (de Villiers et al. 2021), the limited number of cases permits a full factorial design. The eighteen cases appear in random order. The investor can neither skip cases nor navigate back and forth. Having submitted her or his decision, the investor moves to the next case until she or he is finished. The investor adjusts a slider to indicate her or his probability to invest. There is no default probability. The slider is set to zero, but the investor needs to move it to proceed. Thus, if the investor wants to indicate that the probability for her or him to invest in the company is zero, she or he has to move the slider forth and back to zero. If an investor tried to proceed without moving the slider, she or he was told so.

The experiment unfolds in six steps. First, the participant is told that she or he will make a series of investment decisions and that she or he will be given data about the company to invest in for each decision. Second, the investor is introduced to the data, which consist of information about financial performance and up to two more pieces of information: the sustainability agency score and green revenue. Third, the investor is given the example of a company and learns which values each of these three attributes can have. Fourth, the investor takes an attention test consisting of two questions about her or his task. Fifth, the investor proceeds to make, one by one, the eighteen investment decisions in randomized order. Sixth, the investor answers some questions about her or his attitude to the protection of the environment and to government regulation as well as some demographic questions. The experimental instructions are reprinted in Appendix B.

**Dependent and independent variables**

The dependent variable is the probability for the investor to invest in a company (Reimsbach et al. 2018). The participant, in her or his role as an investor, indicates this probability for each instance of a company.

We manipulate the information about the company to invest in, resulting in eighteen cases. First, we are mainly interested in the individual and combined effects of green revenue and sustainability ratings on investment decisions. Green revenue and the sustainability rating can each be either high or low (above or below industry average). Moreover, either may be "not reported," because it may not be mandatory for a company to report green revenue and it is voluntary to report the sustainability rating. Hence, we manipulate either attribute at three levels. Second, we manipulate financial performance at two levels, high or low, to control for its effect. Third, to control for the effect of a pro-environment attitude, we measure this attitude with a set of nine questions, which we adapt from the literature (Ebenbach et al. 1998, Kortenkamp and Moore 2001). Likewise, we adapt another set of questions to measure the pro-government (versus pro-market) attitude (Goff and Noblet 2018). We consider these variables in additional analyses, where we split the sample at the median to distinguish between investors with strong and weak pro-environment attitudes and with strong and weak pro-government attitudes. The questions are reprinted together with the other experimental instructions in Appendix B.

Since the disclosure of green revenue is mandatory for firms that fall under the taxonomy regulation, the case where the company reports green revenue is of primary interest. Whether much or little of the company's revenue is green, it needs to be disclosed. We include the case where green revenue is not reported first because there are companies that are not subject to the regulation. In particular, the regulation applies to European companies only, but investors can naturally invest their money in a European or a comparable non-European company. Second, the case of unreported green revenue allows us to recover the effect of mandatory disclosure in the instance where it hurts—i.e., where the company is required to disclose that little of its revenue is green. The disclosure of a sustainability rating is normally voluntary, and it is therefore unlikely that a company reports a low rating. We include this case for completeness. The three attributes (financial performance, green revenue, and the sustainability rating) are presented to the investor in random order to preclude order effects (Warnick et al. 2018).

**Attention check**

The difference between green revenue and a sustainability rating is subtle. To ensure that our participants understood this difference, we required them to correctly answer two multiple-choice questions. The first question asks which organization determines the sustainability rating (an independent rating agency). The second question asks how green revenue is determined (by the company based on a list of sustainable business activities established by the government). The available answers include incorrect but plausible alternatives. Both the questions and answers figure in the instructions in Appendix B. 99 of our 180 participants (55 percent) answered both questions correctly. The other 81 participants continued to the survey, but we disregard their answers for our analyses. The demographics of the participants who failed the test do not differ significantly from those who passed it. However, the investors who failed the test took only slightly more than eight minutes (with a standard deviation of eight minutes) to finish the survey. This is significantly less time than those who passed the test took ($t = 3.03$, $p = 0.003$).

## IV. RESULTS

**Summary statistics**

We are first interested in the investment probability arising from the disclosure of any combination of much versus little green revenue and a high versus no sustainability rating. Panel A of Table 1 lists the mean investment probability for each of the resulting four conditions along with the averages across conditions. For example, the probability for our 99 investors to invest in a company that reports much green revenue but no sustainability rating averages 50 percent, pooling the high and low levels of financial performance. Figure 1 depicts the means for illustration. We then consider the voluntary disclosure of green revenue, as disclosure is not mandatory for small and non-European companies. Panel B summarizes the investment probabilities for the non-disclosure of green revenue; Panel C, for the disclosure of a low sustainability rating. (A detailed breakdown without averages can be found in Table 6 in Appendix A.) The next three sections report our main results. We then move on to examine how the investment probability depends on investors' attitudes to the protection of the environment and to government regulation. Finally, we study how the financial performance of the company interacts with the disclosure of green revenue and the sustainability rating to determine the investment probability.

—Insert Table 1 about here.—

—Insert Figure 1 about here.—

**Individual effects of green revenue and sustainability rating disclosure**

H1 maintains that the disclosure of much versus little green revenue increases the investment probability. H2 predicts the same effect for the disclosure of a high sustainability rating versus the non-disclosure of any sustainability rating. To test H1 and H2, we regress the investment probability on green revenue, the sustainability rating, and the interaction terms between them. We cluster the errors by investor to account for the dependence of the investment decisions that arise from our within-subjects design. In line with our theory, we choose the disclosure of little green revenue and the non-disclosure of a sustainability rating as baselevels. Panel A of Table 2 reports the regression estimates. The effects predicted by H1 and H2 are obtained by contrast tests. According to Panel A of Table 1, the disclosure of much (rather than little) green revenue increases the investment probability from about 38 to 57 percent. The result of the test in Panel B of Table 2 shows that the increase is significant ($\beta = 18.65$, $t = 14.96$, $p < 0.001$). Likewise, the disclosure of a high sustainability rating, as opposed to the non-disclosure of any sustainability rating, increases the probability by a significant 14 percentage points, from 41 to 55 percent ($\beta = 14.13$, $t = 12.63$, $p < 0.001$). Summing up, the results of our tests support H1 and H2.

—Insert Table 2 about here.—

In Figure 1, the difference between the two lines illustrates H1, whereas the average of the increasing "slopes" reflects H2. In addition to its main effect, the simple effects of the disclosure of green revenue are significant, whether the sustainability rating is unreported or high. Panel A of Table 1 shows that, in the former case, the investment probability increases from 31 to 50 percent ($\beta = 18.26$, $t = 12.32$, $p < 0.001$, as can be seen from Panel A of Table 2); in the latter, from 46 to 64 percent ($\beta = 19.03$, $t = 12.14$, $p < 0.001$, untabulated). The same holds for the simple effects of the disclosure of the sustainability rating, which raises the investment probability from 31 to 46 percent if the company reports little green revenue, according to Panel A of Table 1 ($β = 14.52$, $t = 9.36$, $p < 0.001$, by Panel A of Table 2), and from 50 to 64 percent, if the company reports much green revenue ($β = 13.75$, $t = 10.71$, $p < 0.001$, untabulated). In summary, the increases of the investment probability are about 18 to 19 percentage points for much versus little green revenue, and about 14 to 15 percentage points for a high versus a low sustainability rating.

**Combined effect of green revenue and sustainability rating disclosure**

H3 predicts that the disclosure of a high sustainability rating does not compensate the damage from reporting little green revenue. Indeed, Panel A of Table 1 shows that the investment probability is 46 percent if the company reports little green revenue and a high sustainability rating, but 50, if it reports much green revenue and no sustainability rating. The difference is significant according to Panel B of Table 2 ($β = 4.52$, $t = 3.69$, $p < 0.001$). As expected, investors would rather see one positive signal (i.e., much green revenue) than an ambivalent combination of a positive and a negative signal (little green revenue but a high rating). The former case leaves them uncertain about whether the rating would be high or low if it were disclosed, but it is easier to ignore absent than contradictory information. Taking green revenue as a valid measure of sustainability, they may also infer from much green revenue that the sustainability rating would be high if reported (without necessarily understanding our offsetting argument). The increase of the investment probability from 31 to 50 percent as a result of the disclosure of much rather than little green revenue but no sustainability rating is significant ($β = 18.26$, $t = 12.32$, $p < 0.001$, by Panel A of Table 2), and so is the increase from 31 to 46 percent ($β = 14.52$, $t = 9.36$, $p < 0.001$).

Having considered the combined effect of little green revenue and a high sustainability rating, we turn to H4 about that of good performance on both dimensions. On the one hand, much green revenue normally correlates with a high sustainability rating. Thus, the rating does not convey any new information, which argues against an incremental effect. On the other hand, investors may ignore the empirical correlation and follow a counting heuristic instead. Hence, there are arguments for a negative interaction or none, but not for a positive interaction, which led us to make a weak directional prediction about whether investors discount the sustainability rating. In Figure 1, the "slope" of either of the two lines is almost the same, and hence it does not depend on how much of the company's revenue is green. According to Panel A of Table 1, the investment probability increases by about 14 percentage points in either case (from 31 to 46 and from 50 to 64 percent). Panel A of Table 2 shows that the coefficient on the interaction term is negative but insignificant ($\beta = -0.77$, $t = -0.44$, $p = 0.664$). Hence, we do not find support for any interaction, neither positive nor negative, which confirms H4. Investors count the high rating if it comes in addition to much green revenue, although it would be reasonable to discount it.

**Effect of voluntary versus mandatory disclosure of green revenue**

H5 predicts that the investment probability is higher if the company reports much green revenue against the baseline of non-disclosure. Panel A of Table 1 lists an average investment probability of 57 percent if much revenue is green. According to Panel B, the probability averages 40 percent in the case of non-disclosure, which is higher than the 38 percent if little green revenue is disclosed. In support of H5, the result of the contrast test in Panel B of Table 2 shows that the difference of 17 (= 57 − 40) percentage points between the disclosure of much green revenue and non-disclosure is significant ($\beta = 16.57$, $t = 15.49$, $p < 0.001$). In fact, the increase of the investment probability is significant regardless of whether the company reports a high or no sustainability rating. In the former case, it rises by 15 percentage points, from 49 to 64 percent ($\beta = 14.63$, $t = 11.94$, $p < 0.001$, untabulated); in the latter, by 18 percentage points, from 32 to 50 percent ($\beta = $ 18.51, $t = 12.88$, $p < 0.001$, untabulated). In summary, the effect of the disclosure of much green revenue relative to non-disclosure parallels that of a high sustainability rating considered in H2.

H6 maintains that reporting little green revenue is worse for a company than it would be to not disclose green revenue. According to Panel B of Table 1, the investment probability averages 38 percent in the former case and 40 in the latter. Hence, the disclosure of little green revenue decreases the investment probability by about two percentage points. The result of the contrast test in Panel B of Table 2 shows that this difference is significant ($\beta = -2.07$, $t = -2.41$, $p = 0.017$) and supports H6. The simple effect of non-disclosure is insignificant if the company does not disclose a sustainability rating either ($\beta = -0.52$, $t = -0.51$, $p = 0.612$, by Panel A of Table 2) but significant if it reports a high rating ($\beta = -3.63$, $t = -3.10$, $p = 0.003$, untabulated). In the former case, the investment probability rises from about 31 to 32 percent; in the latter, from 46 to 49, according to Panel B of Table 1. Hence, reporting little green revenue hurts especially a company that has a high sustainability rating and thus sends contradictory signals to investors. For completeness, the disclosure of a low sustainability rating compared to non-disclosure has a similar effect. From Panel C of Table 1, disclosure reduces the probability by more than 4 percentage points, from 41 to 36 percent, and this difference is significant ($\beta = -4.17$, $t = -4.85$, $p < 0.001$, untabulated).

**Effect of investors' pro-environment and pro-government attitudes**

One of the arguments for H1 and H2 is that certain investors are willing to pay a premium for stock from sustainable companies because they have a preference for such companies. These investors should care more about both a company's green revenue and sustainability rating and respond more sensitively to information about these. To capture this preference, we asked the investors in our experiment a series of questions about their pro-environment attitude. The investors indicated their agreement with nine statements on 7-point Likert scales, ranging from 1 to 7. The Cronbach's α of 0.87 indicates a high reliability of the scale, and we average all answer scores into a one-dimensional measure of the investor's pro-environment attitude. For our analyses, we split our sample of investors at the median to distinguish between 49 investors with a weak pro-environment attitude, and 50, with a strong pro-environment attitude. Panel A of Table 3 breaks the investment probability down by pro-environment attitude and condition. Similar to Figure 1, Panel A of Figure 2 depicts the resulting means for illustration. The solid lines connect the mean investment probabilities of investors with a strong pro-environment attitude; the dashed lines, of those with a weak pro-environment attitude.

—Insert Table 3 about here.—

—Insert Figure 2 about here.—

Revisiting H1, the increase of the investment probability if much rather than little green revenue is disclosed should be larger among investors with a strong pro-environment attitude. Indeed, the distance between the midpoints of the solid lines is larger than between those of the dashed lines in Panel A of Figure 2. On average, the investment probability increases by about 23 percentage points (from 35 to 58 percent) among investors with a strong pro-environment attitude, as opposed to 14 (from 42 to 56) among the others in Panel A of Table 3. To test the difference of 9 (= 23 − 14) percentage points for significance, we regress the investment probability on green revenue, sustainability rating, and pro-environment attitude (Panel A of Table 7 in Appendix A). Pro-environment attitude is a dummy variable, which is 1 if the attitude is strong, and 0 else. The contrast test in Panel A of Table 4 shows that the difference is significant ($\beta = 9.19$, $t = 3.86$, $p < 0.001$). Returning to H2, if the company discloses a high rather than no sustainability rating, the investment probability increases by 16 percentage points (from 39 to 55 percent) among investors with a strong pro-environment attitude, and by 12 (from 43 to 55), among the others. The difference between those increases is marginally significant ($\beta = 3.70$, $t = 1.67$, $p = 0.098$).

—Insert Table 4 about here.—

The computation and disclosure of green revenue is required by a government regulation, and it might be perceived differently by investors depending on their attitude to government regulation. We conjecture that investors who approve of government intervention care more about green revenue than investors who trust in the market. To capture our investors' pro-government attitude, we had them indicate their agreement, on 7-point Likert scales, with six statements about government intervention or regulation. With a Cronbach's $\alpha$ of 0.84, the scale turns out reliable, and we average each investor's scores to obtain a measure of her or his pro-government attitude. Following the same procedure as with the pro-environment attitude, we split our sample at the median to discern 51 investors with a strong from 48 with a weak pro-government attitude. Panel B of Table 3 breaks down the investment probability by pro-government attitude and condition. Panel B of Figure 2 plots the means, where the solid lines connect the mean investment probabilities of investors with a strong pro-government attitude; the dashed lines, of investors with a weak pro-government attitude. We note that the pattern of means depicted in Panel B somewhat resembles the pattern for the pro-environment attitude in Panel A of the figure.

The summary statistics in Panel B of Table 3 reveal that the investment probability rises on average by 21 percentage points (from 36 to 57 percent) among investors with a strong pro-government attitude if the company reports much rather than little green revenue. For the other investors, the increase is 16 percentage points (from 41 to 57 percent). We create a dummy variable, which is 1 if the investor's pro-government attitude is strong, and 0 else. Based on a regression of the investment probability on green revenue, sustainability rating, and pro-government attitude, along with any interactions (Panel B of Table 7 in Appendix A), the contrast test in Panel A of Table 4 shows that the difference of close to 7 ($\approx 21 - 16$) percentage points is significant ($\beta = 6.56$, $t = 2.74$, $p = 0.007$). A similar test for whether investors' response to the sustainability rating depends on their pro-government attitude turns out insignificant ($\beta = 0.79$, $t = 0.35$, $p = 0.725$). Hence, although the two attitudes are moderately correlated (Spearman's $\rho = 0.53$, $p < 0.001$), investors' pro-environment attitude drives their response to the disclosure of both much green revenue and a high sustainability rating, whereas their pro-government attitude drives their response to the disclosure of green revenue only, as one would expect.

H3 states that the disclosure of much green revenue beats that of a high sustainability rating. It is straightforward that investors with a strong pro-government attitude rely more on green revenue, relative to the sustainability rating, for their investment decision. The difference is 7 percentage points (= 50 − 43 percent) among these but 1 (= 50 − 49) among the others in Panel B of Table 3. By the second contrast test in Panel B of Table 4, this difference is significant ($\beta = 5.77$, $t = 2.44$, $p = 0.017$). Interestingly, the pro-environment attitude has a similar effect. In Panel B of Table 3, the increase in investment probability resulting from the disclosure of much green revenue exceeds that of the disclosure of a high sustainability rating, combined with little green revenue, by 7 percentage points (= 51 − 44 percent) among investors with a strong pro-environment attitude, but less than 2 ($\approx 49 - 48$) among the others. By the result of the first contrast test listed in Panel B of Table 4, this difference is significant ($\beta = 5.49$, $t = 2.30$, $p = 0.024$). While this observation is expected for the pro-government attitude, it turns out that pro-environment investors are more sensitive to green revenue than to a sustainability rating, too. It seems that, again, green revenue is considered a better measure of sustainability than the rating by them.

**Effect of financial performance**

Our hypotheses do not relate to financial performance. Nonetheless, we manipulated financial performance at two levels in our experiment (high and low) to control for its effect. Although sustainability is crucial, a company's financial performance will normally matter more for the decision to invest in it than its green revenue or sustainability rating. We conjecture that high financial performance is what drives investor's interest in a company. Once investors are interested, they consider the company's green revenue and sustainability rating to adjust their investment probability upward or downward. Hence, we expect that investors respond sensitively to differences in a company's green revenue or sustainability rating if its financial performance is high and they are therefore interested. Conversely, if a company's financial performance is low, investors are less interested in it to begin with, and then they do not care much about its green revenue or sustainability rating. Thus, investors should respond less sensitively to such information in this case. Panel A of Table 5 breaks the investment probability down by financial performance and condition. Figure 3 illustrates the summary statistics. The filled markers depict the means in the case where financial performance is high; the unfilled markers, where it is low.

—Insert Table 5 about here.—

—Insert Figure 3 about here.—

The investment probability is higher if financial performance is high, averaging 73 as opposed to 23 percent from Panel A of Table 5. In response to the disclosure of much rather than little green revenue, the average investment probability rises by 23 percentage points on average if financial performance is high, from 61 to 84 percent, and by 14 if financial performance is low, from 16 to 30 percent. In the figure, the distance between the two lines at the top, which pertain to high financial performance, appears larger than that between the two lines at the bottom, which pertain to low financial performance. To test the difference of 9 (= 23 − 14) percentage points for significance, we regress the investment probability on green revenue, sustainability rating, financial performance, and their interactions, with errors clustered by investor (Table 8 in Appendix A). In Panel B of Table 5, the difference is significant ($\beta = 8.54$, $t = 4.99$, $p < 0.001$). Turning to the rating, the investment probability increases by about 16 percentage points (from 64 to 81 percent) if financial performance is high, and by 12 if it is low (from 17 to 29). In Figure 3, the "slopes" of the lines at the top are steeper than of those at the bottom. Again, the resulting difference of 4 percentage points is significant ($\beta = 3.99$, $t = 2.68$, $p = 0.009$).

Finally, we examine whether the difference between the effects of the disclosure of much green revenue versus a high sustainability rating, which H3 is about, hinges on financial performance, too. If investors are more sensitive to sustainability information once a company's high financial performance has drawn their interest, they are likely more sensitive to the different combinations of such information. Looking at Panel A of Table 5, the disclosure of green revenue raises the investment probability by 7 percentage points (from 69 to 76 percent) if the company's financial performance is high, and by 2 if it is low (from 22 to 24). The result of the contrast test in Panel B confirms that the difference of 5 (= 7 − 2) percentage points is significant ($\beta = 4.55$, $t = 2.36$, $p = 0.020$). This observation offers further support to our argument that investors' sensitivity to environmental information depends on financial performance, being higher if they are more interested in investing in a company to begin with.

## V. CONCLUSION

Sustainability reporting is now considered a means to channel investments into sustainable business and thus mitigate climate change. Governments and authorities around the globe are regulating or consider regulating sustainability reporting. The European Union has moved forward with a taxonomy of sustainable activities and the mandatory disclosure of metrics that are based on it. The rule-based regulation is a step toward standardizing sustainability reporting, and it creates a precedent for regulation and standard setting. In our experiment, we pit green revenue against a third-party assured rating as an alternative sustainability measure. We find, first, that the disclosure of green revenue increases investors' willingness to invest. Second, if a company has little green revenue, it can mitigate but not level out the harm by adding a favorable sustainability rating. Conversely, if it complements the disclosure of much green revenue with a favorable sustainable rating, investors ignore that the two metrics are not normally independent. Third, voluntary disclosure is worthwhile for companies who do not fall under the regulation if they have much but not if they have little green revenue to disclose. These results are obtained in a sample of U.S. residents by a neutral experiment that makes no reference to the European Union.

Our findings hold important managerial implications. Leaving aside the possibility of greenwashing, companies have two ways of making their business more sustainable. They can change their operations to reduce their environmental impact or offset the impact that they make (Johnson et al. 2020). The taxonomy favors operational change over offsetting and thus increases its opportunity cost. Offsetting results most likely in the combination of a favorable sustainability rating with little green revenue. This is because green revenue is not susceptible to offsetting, but sustainability ratings usually are. However, the disclosure of little green revenue reduces the probability for the company to raise funding, even if it is combined with a favorable sustainability rating. Conversely, a favorable sustainability rating normally comes without an extra cost if a company earns much green revenue because its operations are environmentally sustainable. Nonetheless, investors reward the sustainability rating disproportionately in this case. Unsurprisingly, it is always a good idea for a company to report a favorable sustainability rating, whether it has much or little green revenue. That said, it pays surprisingly much off to report a favorable rating in the case where one would think that investors care less for it.

This study contributes to the literature on sustainability reporting, and particularly the regulation of sustainability reporting (Downar et al. 2021, Fiechter et al. 2022). Sustainability reporting has long been mostly voluntary. Mandates, such as the European Union's non-financial disclosure directive, left companies much discretion. Despite high hopes in voluntary disclosure, the evidence is mixed (Clarkson et al. 2013, Dhaliwal et al. 2011, Plumlee et al. 2015). Our results add to the evidence that argues for real effects of reporting mandates. In particular, our experiment enables us to contrast a government regulation with the sustainability rating as a market-based solution. While ratings are popular, they appear to lack standardization and reliability (Christensen et al. 2022). Although we describe the sustainability rating as assured by an independent third party in our experiment, the mandate gives green revenue more pull. Our additional analyses show that investors' reliance on green revenue is influenced by their attitude to government intervention. This observation suggests that investors appreciate the fact that green revenue is a government-backed rule-based metric. That said, our findings are not contingent on the investors' attitude to government regulation and environmental protection but hold in the whole sample.

The European Union's taxonomy regulation has just entered in force and parts of it are still in the making. There is yet uncertainty about how to apply the regulation, which some companies even note in their sustainability reports, let alone about its effects. Our study is supposed to inform field research, not substitute it. We believe that it will be particularly interesting to investigate how companies explain their green revenue and expenditures in their sustainability reports and how they complement these metrics with sustainability ratings. Our findings suggest that taxonomy-aligned reporting will not crowd out sustainability ratings. Instead, companies that have little green revenue to disclose will experience heightened pressure to resort to sustainability ratings as alternative proof of their sustainability to avert damage, and to explain why ratings are better metrics to measure their sustainability. Companies that have much green revenue, in turn, will likely rely on sustainability ratings to further stress the sustainability of their business. Talking about the relationship between reporting and the underlying performance, it will be exciting to see whether the combination of little green revenue and a high rating (i.e., incongruent information) correlates with the use of offsetting, which companies typically disclose in their reports.

# REFERENCES

Al-Tuwaijri, S. A., Christensen, T. E., & Hughes, K. E. (2004). The relations among environmental disclosure, environmental performance, and economic performance: A simultaneous equations approach. *Accounting, Organizations and Society, 29*(5/6), 447–471. https://doi.org/10.1016/S0361-3682(03)00032-1

Arribas, I., Espinós-Vañó, M. D., García, F., & Morales-Bañuelos, P. B. (2019). The inclusion of socially irresponsible companies in sustainable stock indices. *Sustainability, 11*(7), 2047. http://doi.org/10.3390/su11072047

Berinsky, A. J., Huber, G. A., & Lenz, G. S. (2012). Evaluating online labor markets for experimental research: Amazon.com's Mechanical Turk. *Political Analysis, 20*(3), 351–368. https://doi.org/10.1093/pan/mpr057

Buchheit, S., Doxey, M. M., Pollard, T., & Stinson, S. R. (2017). A technical guide to using Amazon's Mechanical Turk in behavioral accounting research. *Behavioral Research in Accounting, 30*(1): 111–122. https://doi.org/10.2308/bria-51977

Buhrmester, M., Kwang, T., & Gosling, S. D. (2011). Amazon's Mechanical Turk: A new source of inexpensive, yet high-quality, data? *Perspectives on Psychological Science, 6*(1), 3–5. https://doi.org/10.1177/1745691610393980

Casler, K., Bickel, L., & Hackett, E. (2013). Separate but equal? A comparison of participants and data gathered via Amazon's MTurk, social media, and face-to-face behavioral testing. *Computers in Human Behavior, 29*(6), 2156–2160. https://doi.org/10.1016/j.chb.2013.05.009

Cheng, B., Ioannou, I., & Serafeim, G. (2014). Corporate social responsibility and access to finance. *Strategic Management Journal, 35*(1), 1–23. https://doi.org/10.1002/smj.2131

Christensen, D. M., Serafeim, G., & Sikochi, A. (2022). Why is corporate virtue in the eye of the beholder? The case of ESG ratings. *The Accounting Review, 97*(1), 147–175. https://doi.org/10.2308/TAR-2019-0506

Clarkson, P. M., Fang, X., Li, Y., & Richardson, G. (2013). The relevance of environmental disclosures: Are such disclosures incrementally informative? *Journal of Accounting and Public Policy, 32*(5), 410–431. https://doi.org/10.1016/j.jaccpubpol.2013.06.008

Continental (2022). Annual report 2021. https://www.continental.com/en/investors/reports/

Crump, M. J., McDonnell, J. V., & Gureckis, T. M. (2013). Evaluating Amazon's Mechanical Turk as a tool for experimental behavioral research. *PloS ONE, 8*(3), e57410. https://doi.org/10.1371/journal.pone.0057410

De Villiers, C., Cho, C. H., Turner, M. J., & Scarpa, R. (2021). Are shareholders willing to pay for financial, social and environmental disclosure? A choice-based experiment. *European Accounting Review*. https://doi.org/10.1080/09638180.2021.1944890

Dhaliwal, D. S., Li, O. Z., Tsang, A., & Yang, Y. G. (2011). Voluntary nonfinancial disclosure and the cost of equity capital: The initiation of corporate social responsibility reporting. *The Accounting Review, 86*(1), 59–100. https://doi.org/10.2308/accr.00000005

Dhanda, K. K., & Hartman, L. P. (2011). The ethics of carbon neutrality: A critical examination of voluntary carbon offset providers. *Journal of Business Ethics*, *100*(1), 119–149. https://doi.org/10.1007/s10551-011-0766-4

Delegated Regulation 2021/2139. European Commission. http://data.europa.eu/eli/reg_del/2021/2139/oj

Directive 2014/95/EU. European Parliament, Council of the European Union. http://data.europa.eu/eli/dir/2014/95/oj.

Donofrio, S., Maguire, P., Myers, K., Daley, C., & Lin, K. (2021). Markets in motion. State of the voluntary carbon markets 2021. Installment 1. Forest Trends Association. https://www.forest-trends.org/publications/state-of-the-voluntary-carbon-markets-2021/

Downar, B., Ernstberger, J., Reichelstein, S., Schwenen, S., & Zaklan, A. (2021). The impact of carbon disclosure mandates on emissions and financial operating performance. *Review of Accounting Studies 26*, 1137–1175. https://doi.org/10.1007/s11142-021-09611-x

Ebenbach, D. H., Moore, C. F., & Parsil, S. A. (1998). Internally and externally motivated environmental attitudes. Paper presented at the Annual Convention of the Midwestern Psychological Association, Chicago, IL.

Egginton, J. F., & McBrayer, G. A. (2019). Does it pay to be forthcoming? Evidence from CSR disclosure and equity market liquidity. *Corporate Social Responsibility and Environmental Management, 26*(2), 396–407. https://doi.org/10.1002/csr.1691

El Ghoul, S., Guedhami, O., Kwok, C. C., & Mishra, D. R. (2011). Does corporate social responsibility affect the cost of capital? *Journal of Banking & Finance, 35*(9), 2388–2406. https://doi.org/10.1016/j.jbankfin.2011.02.007

Elliott, W. B., Jackson, K. E., Peecher, M. E., & White, B. J. (2014). The unintended effect of corporate social responsibility performance on investors' estimates of fundamental value. *The Accounting Review, 89*(1), 275–302. https://doi.org/10.2308/accr-50577

European Commission (2017). Guidelines on non-financial reporting. https://eur-lex.europa.eu/legal-content/TXT/?uri=CELEX:52017XC0705(01)

European Commission (2018). A European strategic long-term vision for a prosperous, modern, competitive and climate neutral economy. https://eur-lex.europa.eu/legal-content/TXT/?uri=CELEX:52018DC0773

European Commission (2021). Proposal for a directive of the European Parliament and of the Council amending Directive 2013/34/EU, Directive 2004/109/EC, Directive 2006/43/EC and Regulation (EU) No 537/2014, as regards corporate sustainability reporting. https://eur-lex.europa.eu/legal-content/EN/TXT/?uri=CELEX:52021PC0189

Fiechter, P., Hitz, J.-M., & Lehmann, N. (2022). Real effects of a widespread CSR reporting mandate: Evidence from the European Union's CSR Directive. *Journal of Accounting Research*. https://doi.org/10.1111/1475-679X.12424

Goff, S. H., & Noblet, C. L. (2018). Efficient, but immoral? Assessing market attitudes as multidimensional. *Economics Letters, 170*, 96–99. https://doi.org/10.1016/j.econlet.2018.05.020

Guiral, A., Moon, D., Tan, H. T., & Yu, Y. (2020). What drives investor response to CSR performance reports? *Contemporary Accounting Research, 37*(1), 101–130. https://doi.org/10.1111/1911-3846.12521

Holm, C., & Rikhardsson, P. (2008). Experienced and novice investors: Does environmental information influence investment allocation decisions? *European Accounting Review, 17*(3), 537–557, https://doi.org/10.1080/09638180802016627

Horton, J. J., Rand, D. G., & Zeckhauser, R. J. (2011). The online laboratory: Conducting experiments in a real labor market. *Experimental Economics, 14*, 399–425. https://doi.org/10.1007/s10683-011-9273-9

Johnson, J. A., Theis, J., Vitalis, A., & Young, D. (2020). The influence of firms' emissions management strategy disclosures on investors' valuation judgments. *Contemporary Accounting Research, 37*(2), 642–664. https://doi.org/10.1111/1911-3846.12545

Koonce, L., & Lipe, M. G. (2017). Firms with inconsistently signed earnings surprises: Do potential investors use a counting heuristic? *Contemporary Accounting Research, 34*(1), 292–313. https://doi.org/10.1111/1911-3846.12235

Kortenkamp, K. V., & Moore, C. F. (2001). Ecocentrism and anthropocentrism: Moral reasoning about ecological commons dilemmas. *Journal of Environmental Psychology, 21*(3), 261–272. https://doi.org/10.1006/jevp.2001.0205

Matsumura, E. M., Prakash, R., & Vera-Munoz, S. C. (2014). Firm-value effects of carbon emissions and carbon disclosures. *The Accounting Review, 89*(2), 695–724. https://doi.org/10.2308/accr-50629

Paolacci, G., Chandler, J., & Ipeirotis, P. G. (2010). Running experiments on Amazon Mechanical turk. *Judgment and Decision Making, 5*(5), 411–419. https://doi.org/10.1037/t69659-000

Platform on Sustainable Finance (2021, August 3). Draft report by the Platform on Sustainable Finance on preliminary recommendations for technical screening criteria for the EU taxonomy https://ec.europa.eu/info/publications/210803-sustainable-finance-platform-technical-screening-criteria-taxonomy-report_en

Plumlee, M., Brown, D., Hayes, R. M., & Marshall, R. S. (2015). Voluntary environmental disclosure quality and firm value: Further evidence. *Journal of Accounting and Public Policy, 34*(4), 336–361. http://doi.org/10.1016/j.jaccpubpol.2015.04.004

Regulation 2020/852. European Parliament, Council of the European Union. http://data.europa.eu/eli/reg/2020/852/oj

Reimsbach, D., Hahn, R., & Gürtürk, A. (2018). Integrated reporting and assurance of sustainability information: An experimental study on professional investors' information processing. *European Accounting Review, 27*(3), 559–581. https://doi.org/10.1080/09638180.2016.1273787

Richardson, A. J., Welker, M., & Hutchinson, I. R. (1999). Managing capital market reactions to corporate social resposibility. *International Journal of Management Reviews, 1*(1), 17–43. https://doi.org/10.1111/1468-2370.00003

Richardson, A. J., & Welker, M. (2001). Social disclosure, financial disclosure and the cost of equity capital. *Accounting, Organizations and Society, 26*(7/8), 597–616. https://doi.org/10.1016/S0361-3682(01)00025-3

Scalet, S., & Kelly, T. F. (2010). CSR rating agencies: What is their global impact? *Journal of Business Ethics, 94*(1), 69–88. https://doi.org/10.1007/s10551-009-0250-6

SEC (2010, February 8). Commission guidance regarding disclosure related to climate change. https://www.sec.gov/rules/interp/2010/33-9106.pdf

SEC (2022, March 21). The enhancement and standardization of climate-related disclosures for investors. https://www.sec.gov/rules/proposed/2022/33-11042.pdf

Simnett, R., Vanstraelen, A., & Chua, W. F. (2009). Assurance on sustainability reports: An international comparison. *The Accounting Review, 84*(3), 937–967. https://doi.org/10.2308/accr.2009.84.3.937

Tang, M., & Venkataraman, S. (2018). How patterns of past guidance provision affect investor judgments: The joint effect of guidance frequency and guidance pattern consistency. *The Accounting Review, 93*(3), 327–348. https://doi.org/10.2308/accr-51905

Ungeheuer, M., & Weber, M. (2021). The perception of dependence, investment decisions, and stock prices. *The Journal of Finance, 76*(2), 797–844. https://doi.org/10.1111/jofi.12993

Warnick, B. J., Murnieks, C. Y., McMullen, J. S., & Brooks, W. T. (2018). Passion for entrepreneurship or passion for the product? A conjoint analysis of angel and VC decision-making. *Journal of Business Venturing, 33*(3), 315–332. https://doi.org/10.1016/j.jbusvent.2018.01.002

Windolph, S. E. (2011). Assessing corporate sustainability through ratings: Challenges and their causes. *Journal of Environmental Sustainability, 1*(1), 37–57. https://doi.org/10.14448/jes.01.0005

# APPENDIX A

Table 6 shows detailed breakdowns of the investment probability.

—Insert Table 6 about here.—

To complement the results reported in Table 4 and Panels B and C of Table 5, we include Tables 7 and 8. Table 7 reports the results of the regressions underlying the contrasts in Table 4; Table 8, the results of the regression underlying the contrasts in Table 5.

—Insert Table 7 about here.—

—Insert Table 8 about here.—

**APPENDIX B**

**Instructions**

Thank you for agreeing to participate! In this study, we ask you to assume that you are a **private investor in search of an investment opportunity.** You will be asked to make a series of **investment decisions.** For each decision, you will be given **information about a company.**

The study consists of **three parts:**

1. Introduction to the types of information about the companies;
2. Eighteen investment decisions;
3. Further questions.

**Description of company data**

Your company data will include one to three pieces of information: financial performance, the Sustainability Agencycscore, and green revenue.

*Financial performance*

Financial performance refers to your potential gains from holding shares of the company, both from dividend payments and capital appreciation (increase in company value).

*Sustainability Agency score*

Sustainability Agency is an **independent non-profit organization** that **scores companies for sustainability.** Companies that want to be scored report to the Sustainability Agency, which publishes the scores online. **Those companies can include their score in their annual report.** The Sustainability Agency provides those scores to help investors make better investment decisions with respect to **climate change and other environmental impacts.** The Sustainability Agency's scores are based on information from questionnaires, which are updated regularly. Scoring criteria include the quantity and quality of environmental information disclosed as well as the management's actions to manage the company's environmental impact.

*Green revenue*

**The government** has adopted a **list of environmentally sustainable business activities,** the "taxonomy." Companies who fall under the taxonomy law must match their business activities against that list. **These companies label revenue from listed activities as "green" in their annual reports.** The purpose of the law is to help investors make better investment decisions with respect to **climate change and other environmental impacts.** For a business activity to be green, it must meet technical criteria such as low $CO_2$ emissions, which are updated regularly. Moreover, the business activity must comply with social minimum safeguards (e.g., workplace safety).

**Example**

Please find below an example for a company you are going to evaluate hereafter.
Company data:

| Financial Performance | Low |
| Sustainability Agency score | High |
| Percentage of green revenue | High |

Low denotes below industry average. High denotes above industry average.

Companies can voluntarily choose not to report their Sustainability Agency score and might not be obligated to report the percentage of green revenue, or both. In these cases, the table will say "not reported."

**Attention Checks**

Please answer the following questions.[6]

Which organization determines the company's sustainability score?

- A governmental agency.
- An independent rating agency.
- A group of private businesses.
- An international consortium of banks.

How is green revenue determined?

- By a rating agency based on publicly available information.
- By an auditor based on a list of sustainable business activities.
- By the company based on a list of sustainable business activities established by the government.
- By a rating agency based on questionnaires to be filled in by the company.

**Task**

You will assess **eighteen companies from the aluminum industry.** Some of these companies report their Sustainability Agency score, some report their green revenue, some both, some neither.

If a company does not report a Sustainability Agency score or its green revenue, you **cannot infer that this company is hiding information.** Sustainability Agency scoring is voluntary, and the company may just not fall under the taxonomy law.

The aluminum industry is considered an **energy-intensive industry with high $CO_2$ emission levels.**

**Please indicate the probability that you would invest in each of the companies.**

**Company**[7]

Company data:

| Financial Performance | High |
|---|---|
| Sustainability Agency score | Low |
| Percentage of green revenue | Low |

What is the probability that you would invest in this firm?[8]

---

[6] The order of the answer options is randomized. Correct answers : "An independent rating agency." "By the company based on a list of sustainable business activities …"

[7] The order of both the cases and the attributes within in each case are randomized.

[8] A slider is adjusted to indicate the percentage. The slider is set to 0 percent by default but must be moved, even if the investor's answer is "0 percent."

```
                                    0  10  20  30  40  50  60  70  80  90  100
Investment probability (from 0% to 100%)  ●————————————————●
```

**Questions**[9]

Please indicate your agreement with each of the following statements.
1. I try hard to carry my pro-environmental beliefs over into all the other parts of my life.
2. Because of my personal values, I believe that ignoring environmental matters is OK.
3. When it comes to questions about the environment, I feel driven to know the truth.
4. According to my personal values, ignoring human impacts on the larger ecosystem is OK.
5. I am motivated by my personal beliefs to try to protect the environment.
6. The interrelatedness of all living things in the ecosystem is something I have never felt personally compelled to consider.
7. What happens to the larger ecosystem, beyond what happens to humans, doesn't make much difference to me.
8. I have not found it essential to try to protect the larger ecosystem, beyond what happens to humans.
9. It is personally important to me to try to protect the larger ecosystem, beyond what happens to humans.

**Questions (continued)**[10]

Please indicate your agreement with each of the following statements.
1. In my opinion, it is never acceptable for the government to intervene in markets.
2. In my opinion, the market rules and regulations the government sets are necessary to protect citizens and the environment.
3. In my opinion, government regulation of business usually does more harm than good.
4. In my opinion, markets dominated by only one or a few businesses should be regulated by the government.
5. In my opinion, market systems require a lot of government control to be efficient.
6. In my opinion, there are some goods and services which should not be exchanged through a free market system.
7. Whom do you trust more to cope with the challenges of climate change: government or non-governmental organizations (e.g., business firms, non-profit organizations, and the like)?

---

[9] The statements are rated on a 7-point Likert scale ranging from "Fully disagree" (1) to "Fully agree" (7). The items 2, 4, and 6–8 are reversely coded.

[10] The statements (1–6) are rated on a 7-point Likert scale ranging from "Fully disagree" (1) to "Fully agree" (7). The question (7) is answered on a 7-point scale, too, with the anchors "Government" (1) to "Non-governmental organizations" (7). The items 2 and 4–7 are reversely coded.

# FIGURE 1
## Summary Statistics

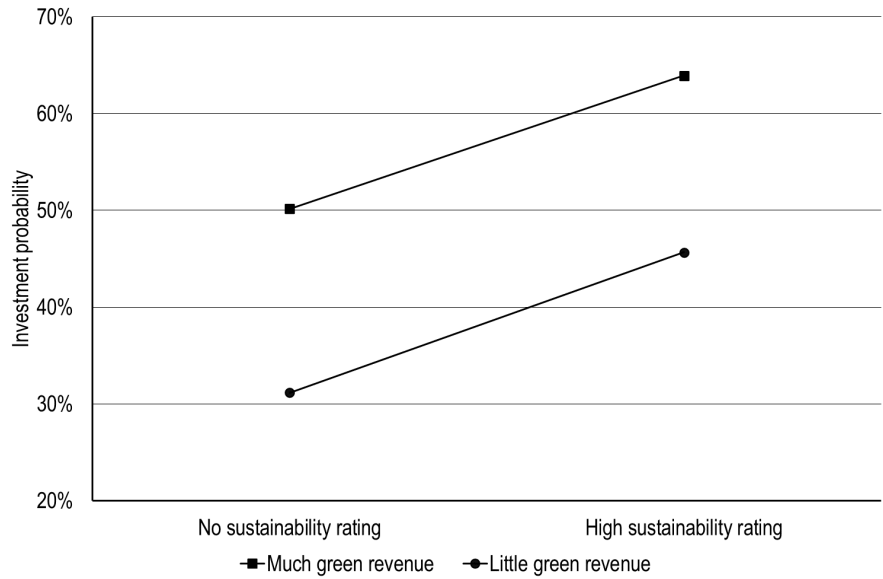

Investment probability depending on green revenue (square versus disk markers) and the sustainability rating. The graph illustrates the relevant summary statistics from Table 1.

**FIGURE 2**
**Pro-Environment and Pro-Government Attitudes**

**Panel A: Pro-environment attitude**

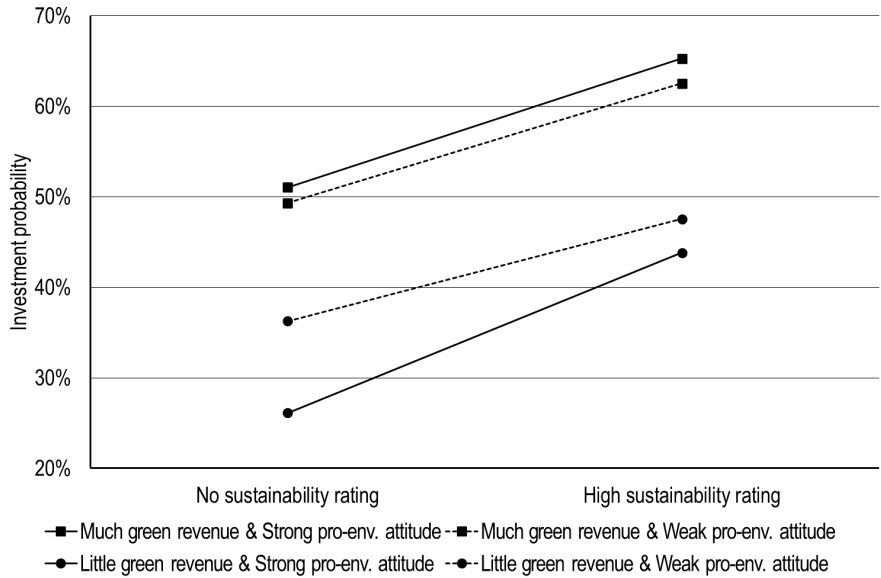

**Panel B: Pro-government attitude**

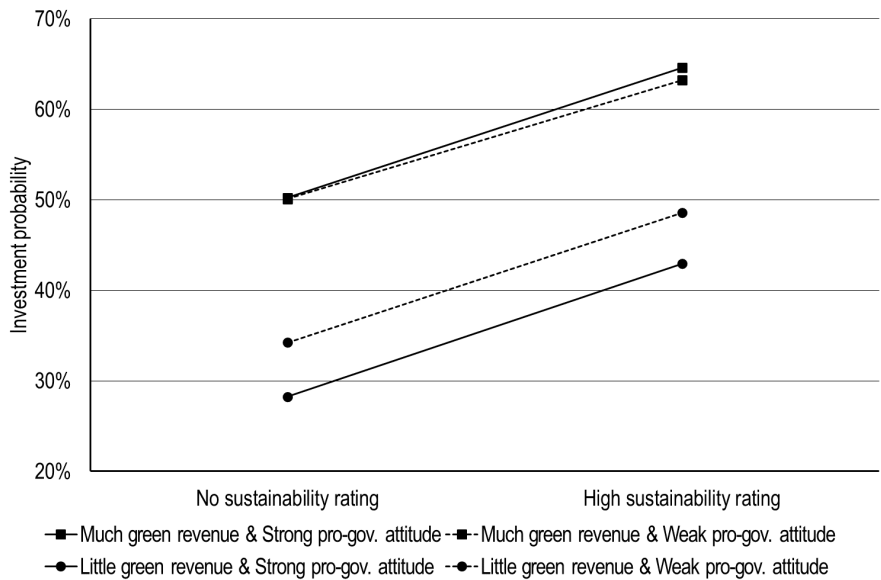

Investment probability depending on green revenue (square versus disk markers), sustainability rating, and pro-environment attitude (Panel A) or pro-government attitude (Panel B, solid versus dashed lines). The graph illustrates the summary statistics from Panels A and B of Table 3.

# FIGURE 3
## Financial Performance

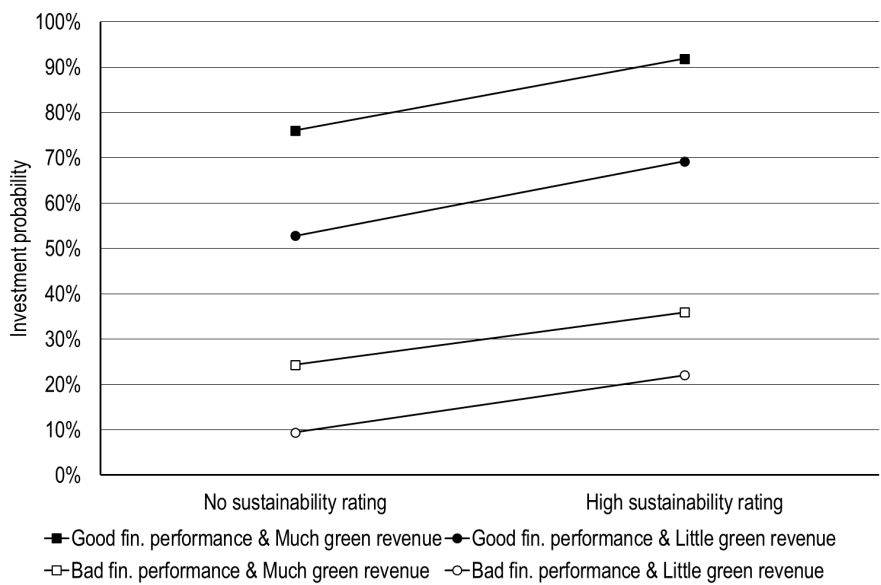

Investment probability depending on financial performance (filled versus unfilled makers, high versus low performance) and green revenue (square versus disk makers). The graph illustrates the summary statistics from Panel A of Table 5.

## TABLE 1
## Summary Statistics

**Panel A: Much green revenue and high sustainability rating with baselines**

|  | Green revenue | | |
|---|---|---|---|
| Sustainability rating | Little | Much | Total |
| Unreported | 31.15 (29.19) | 50.18 (32.70) | 40.67 (32.39) |
| High | 45.67 (32.00) | 63.93 (35.23) | 54.80 (34.83) |
| Total | 38.41 (31.44) | 57.06 (34.64) | 47.73 (34.35) |

**Panel B: No green revenue and high sustainability rating with baselines**

|  | Green revenue | | |
|---|---|---|---|
| Sustainability rating | Little | Unreported | Total |
| Unreported | 31.15 (29.19) | 31.67 (29.54) | 31.41 (29.33) |
| High | 45.67 (32.00) | 49.30 (32.61) | 47.48 (32.32) |
| Total | 38.41 (31.44) | 40.48 (32.30) | 39.45 (31.87) |

**Panel C: Much green revenue and no sustainability rating with baselines**

|  | Green revenue | | |
|---|---|---|---|
| Sustainability rating | Little | Much | Total |
| Low | 27.41 (28.19) | 45.58 (31.19) | 36.49 (31.05) |
| Unreported | 31.15 (29.19) | 50.18 (32.70) | 40.67 (32.39) |
| Total | 29.28 (28.72) | 47.88 (32.00) | 38.58 (31.78) |

Mean and, in parentheses, standard deviation of the investment probability (in percent) by condition. The number of observations is 99 throughout. For a detailed breakdown (without row and column totals) see Table 6 in Appendix A.

## TABLE 2
## Hypothesis Tests

**Panel A: Regression to test Hypotheses 1–6**

|  | Coefficient | Standard error | $p$-value |
|---|---:|---:|---:|
| Green revenue | | | |
|     Unreported | 0.52 | 1.01 | 0.612 |
|     Much | 19.03 | 1.57 | < 0.001 |
| Sustainability rating | | | |
|     Low | −3.74 | 1.04 | 0.001 |
|     High | 14.52 | 1.55 | < 0.001 |
| Green revenue × Sustainability rating | | | |
|     Unreported × Low | 2.83 | 1.45 | 0.054 |
|     Unreported × High | 3.12 | 1.37 | 0.025 |
|     High × Low | −0.86 | 1.52 | 0.572 |
|     High × High | −0.77 | 1.76 | 0.664 |
| Constant | 31.15 | 1.45 | < 0.001 |

**Panel B: Tests of Hypotheses 1–6**

|  | Coefficient | Standard error | $p$-value |
|---|---:|---:|---:|
| Much vs. little green revenue (H1) | 18.65 | 1.25 | < 0.001 |
| High vs. no sustainability rating (H2) | 14.13 | 1.12 | < 0.001 |
| Much green revenue vs. high sustainability rating (H3) | 4.52 | 1.22 | < 0.001 |
| Much green revenue and high sustainability rating (H4) | −0.77 | 1.76 | 0.664 |
| Much vs. no green revenue (H5) | 16.57 | 1.07 | < 0.001 |
| No vs. little green revenue (H6) | −2.07 | 0.86 | 0.017 |

Panel A: Results of an ordinary least squares regression with errors clustered by investor (1,782 observations = 99 clusters × 18 observations per cluster, $R^2 = 0.12$). The dependent variable is the investor's probability of investment. Green revenue has three levels: much, little, and unreported, where little is the baselevel. The sustainability rating has three levels: high, low, and unreported, where unreported is the baselevel.

Panel B: Tests of the hypothesized effects. The coefficient on the interaction term to test H4 figures among the results of the regression. The other effects result from planned contrast tests.

# TABLE 3
# Pro-Environment and Pro-Government Attitudes

**Panel A: Pro-environment attitude**

| Sustainability rating | Weak pro-environment attitude | | | Strong pro-environment attitude | | |
|---|---|---|---|---|---|---|
| | Green revenue | | | Green revenue | | |
| | Little | Much | Total | Little | Much | Total |
| Unreported | 36.28 (31.29) | 49.31 (32.82) | 42.79 (32.64) | 26.13 (26.18) | 51.04 (32.72) | 38.58 (32.09) |
| High | 47.56 (34.18) | 62.54 (36.15) | 55.05 (35.89) | 43.81 (29.76) | 65.29 (34.43) | 54.55 (33.86) |
| Total | 41.92 (33.17) | 55.92 (35.07) | 48.92 (34.80) | 34.97 (29.33) | 58.17 (34.26) | 46.57 (33.90) |

**Panel B: Pro-government attitude**

| Sustainability rating | Weak pro-government attitude | | | Strong pro-government attitude | | |
|---|---|---|---|---|---|---|
| | Green revenue | | | Green revenue | | |
| | Little | Much | Total | Little | Much | Total |
| Unreported | 34.24 (30.62) | 50.11 (33.33) | 42.18 (32.90) | 28.25 (27.62) | 50.25 (32.25) | 39.25 (31.92) |
| High | 48.57 (33.92) | 63.23 (35.92) | 55.90 (35.61) | 43.93 (29.99) | 64.59 (34.74) | 53.76 (34.14) |
| Total | 41.41 (33.02) | 56.67 (35.18) | 49.04 (34.92) | 35.59 (29.69) | 57.42 (34.20) | 46.50 (33.80) |

Panel A: Mean and, in parentheses, standard deviation of the investment probability by pro-environment attitude and condition. The sample is split at the median into 49 investors with weak pro-environment attitude and 50 investors with strong pro-environment attitude. The pro-environment attitude averages the answers, on 7-point Likert scales, to nine questions (Cronbach's $\alpha = 0.87$).

Panel B: Mean and, in parentheses, standard deviation of the investment probability by pro-government attitude and condition. The median split discerns 48 investors with a weak pro-government attitude; 51, with a strong pro-government attitude. The pro-government attitude averages the answers to six questions on 7-point Likert scales (Cronbach's $\alpha = 0.84$).

The questions about the pro-environment and pro-government attitudes figure in the instructions, which are reprinted in Appendix B.

## TABLE 4
## Interaction with Pro-Environment and Pro-Government Attitudes

**Panel A: Interaction between green revenue, sustainability rating, and attitudes**

|  | Coefficient | Standard error | $p$-value |
|---|---|---|---|
| Green revenue × Pro-environment attitude | 9.19 | 2.32 | < 0.001 |
| × Pro-government attitude | 6.56 | 2.40 | 0.007 |
| Sustainability rating × Pro-environment attitude | 3.70 | 2.22 | 0.098 |
| × Pro-government attitude | 0.79 | 2.24 | 0.725 |

**Panel B: Interaction between green revenue vs. the sustainability rating and attitudes**

|  | Coefficient | Standard error | $p$-value |
|---|---|---|---|
| (Much green revenue − High sustainability rating) |  |  |  |
| × Pro-environment attitude | 5.49 | 2.39 | 0.024 |
| × Pro-government attitude | 5.77 | 2.37 | 0.017 |

---

Planned contrasts, based on ordinary least squares regressions, with clustered errors, of the investment probability on green revenue, the sustainability rating, and either the pro-environment or the pro-government attitude. The results of the regressions are found in Table 6 in Appendix A.

Panel A: Interaction of the effect of green revenue (H1) and the sustainability rating (H2) with the pro-environment and pro-government attitudes. Panel B: Interaction of the difference between the effects of green revenue and the sustainability rating with the pro-environment and pro-government attitudes.

# TABLE 5
## Interaction with Financial Performance

**Panel A: Summary statistics**

| Sustainability rating | Low financial performance | | | High financial performance | | |
|---|---|---|---|---|---|---|
| | Green revenue | | | Green revenue | | |
| | Little | Much | Total | Little | Much | Total |
| Unreported | 9.45 (14.33) | 24.34 (21.40) | 16.90 (19.64) | 52.85 (23.60) | 76.02 (18.50) | 64.43 (24.13) |
| High | 22.10 (21.70) | 35.97 (26.63) | 29.04 (25.21) | 69.23 (21.57) | 91.89 (14.38) | 80.56 (21.53) |
| Total | 15.78 (19.40) | 30.16 (24.79) | 22.97 (23.37) | 61.04 (24.00) | 83.95 (18.34) | 72.50 (24.22) |

**Panel B: Interaction between green revenue, sustainability rating, and financial performance**

| | Coefficient | Standard error | $p$-value |
|---|---|---|---|
| Green revenue × Financial performance | 8.54 | 1.71 | < 0.001 |
| Sustainability rating × Financial performance | 3.99 | 1.49 | 0.009 |

**Panel C: Interaction between green revenue vs. the sustainability rating and financial performance**

| | Coefficient | Standard error | $p$-value |
|---|---|---|---|
| (Much green revenue − High sustainability rating) × Financial performance | 4.55 | 1.92 | 0.020 |

Panel A: Mean and, in parentheses, standard deviation of the investment probability (in percent) by financial performance and condition. The number of observations is 99 throughout.

Panels B and C: Contrasts, based on an ordinary least squares regression, with clustered errors, of the investment probability on green revenue, the sustainability rating, and financial performance. The results of the regressions are found in Table 8 in Appendix A.

Panel A shows the interaction of the effect of green revenue (H1) and the sustainability rating (H2) with financial performance. Panel B shows the interaction of the difference between the effects of green revenue and the sustainability rating (H3) with financial performance.

**TABLE 6**
**Detailed Summary Statistics**

**Panel A: Green revenue and sustainability rating**

| Sustainability rating | Green revenue | | |
|---|---|---|---|
| | Little | Unreported | Much |
| Low | 27.41 (28.19) | 30.75 (28.93) | 45.58 (31.19) |
| Unreported | 31.15 (29.19) | 31.67 (29.54) | 50.18 (32.70) |
| High | 45.67 (32.00) | 49.30 (32.61) | 63.93 (35.23) |

**Panel B: Green revenue, sustainability rating, and financial performance**

| Sustainability rating | Low financial performance | | | High financial performance | | |
|---|---|---|---|---|---|---|
| | Green revenue | | | Green revenue | | |
| | Little | Unreported | Much | Little | Unreported | Much |
| Low | 5.71 (10.58) | 11.04 (17.93) | 22.41 (19.81) | 49.11 (23.10) | 50.46 (24.00) | 68.74 (21.90) |
| Unreported | 9.45 (14.33) | 9.44 (14.95) | 24.34 (21.40) | 52.85 (23.60) | 53.89 (23.08) | 76.02 (18.50) |
| High | 22.10 (21.70) | 24.37 (21.58) | 35.97 (26.63) | 69.23 (21.57) | 74.22 (20.41) | 91.89 (14.38) |

Mean and, in parentheses, standard deviation, of the investment probability (in percent) by condition (Panel A) and by financial performance and condition (Panel B). The number of observations is 99 throughout.

# TABLE 7
# Interaction with Pro-Environment and Pro-Government Attitudes

**Panel A: Pro-environment attitude**

|  | Coefficient | Standard error | p-value |
|---|---:|---:|---:|
| Green revenue | | | |
|     Unreported | −1.46 | 1.36 | 0.287 |
|     Much | 13.03 | 1.92 | < 0.001 |
| Sustainability rating | | | |
|     Low | −4.68 | 1.81 | 0.011 |
|     High | 11.29 | 2.16 | < 0.001 |
| Pro-environment attitude | −10.15 | 2.71 | < 0.001 |
| Green revenue × Sustainability rating | | | |
|     Unreported × Low | 5.74 | 2.25 | 0.012 |
|     Unreported × High | 1.19 | 2.00 | 0.551 |
|     Much × Low | 2.00 | 2.08 | 0.343 |
|     Much × High | 1.95 | 2.52 | 0.441 |
| Green revenue × Pro-environment attitude | | | |
|     Unreported × Strong | 3.91 | 1.99 | 0.052 |
|     Much × Strong | 11.88 | 2.90 | < 0.001 |
| Sustainability rating × Pro-environment attitude | | | |
|     Low × Strong | 1.86 | 2.09 | 0.374 |
|     High × Strong | 6.39 | 3.04 | 0.038 |
| Green revenue × Sustainability rating × Pro-env. attitude | | | |
|     Unreported × Low × Strong | −5.77 | 2.85 | 0.046 |
|     Unreported × High × Strong | 3.81 | 2.72 | 0.165 |
|     Much × Low × Strong | −5.63 | 2.99 | 0.063 |
|     Much × High × Strong | −5.38 | 3.49 | 0.126 |
| Constant | 36.28 | 2.07 | < 0.001 |

**TABLE 7 (continued)**

**Panel B: Pro-government attitude**

|  | Coefficient | Standard error | p-value |
|---|---:|---:|---:|
| Green revenue | | | |
|     Unreported | −1.44 | 1.67 | 0.393 |
|     Much | 15.88 | 2.15 | < 0.001 |
| Sustainability rating | | | |
|     Low | −3.48 | 1.64 | 0.037 |
|     High | 14.33 | 2.11 | < 0.001 |
| Pro-government attitude | −5.99 | 2.82 | 0.036 |
| Green revenue × Sustainability rating | | | |
|     Unreported × Low | 3.50 | 2.22 | 0.118 |
|     Unreported × High | 0.50 | 2.05 | 0.808 |
|     Much × Low | −0.21 | 1.98 | 0.916 |
|     Much × High | −1.22 | 2.66 | 0.648 |
| Green revenue × Pro-government attitude | | | |
|     Unreported × Strong | 3.79 | 2.02 | 0.063 |
|     Much × Strong | 6.13 | 3.08 | 0.049 |
| Sustainability rating × Pro-government attitude | | | |
|     Low × Strong | −0.51 | 2.10 | 0.808 |
|     High × Strong | 0.35 | 3.10 | 0.910 |
| Green revenue × Sustainability rating × Pro-gov. attitude | | | |
|     Unreported × Low × Strong | −1.30 | 2.92 | 0.656 |
|     Unreported × High × Strong | 5.08 | 2.71 | 0.063 |
|     Much × Low × Strong | −1.27 | 3.03 | 0.676 |
|     Much × High × Strong | 0.88 | 3.54 | 0.805 |
| Constant | 34.24 | 1.88 | < 0.001 |

Panel A: Results of an ordinary least squares regression, with errors clustered by investor, of the investment probability on green revenue, sustainability rating, and pro-environment attitude (observations as in Table 2, $R^2 = 0.13$). By median split, pro-environment attitude has two levels: strong and weak.

Panel A: Results of an ordinary least squares regression, with errors clustered by investor, of the investment probability on green revenue, sustainability rating, and pro-government attitude (observations as in Table 2, $R^2 = 0.12$). By median split, pro-government attitude has two levels: strong and weak.

# TABLE 8
# Interaction with Financial Performance

|  | Coefficient | Standard error | $p$-value |
|---|---|---|---|
| Green revenue |  |  |  |
|    Unreported | −0.01 | 1.03 | 0.992 |
|    Much | 14.89 | 1.70 | < 0.001 |
| Sustainability rating |  |  |  |
|    Low | −3.75 | 0.95 | < 0.001 |
|    High | 12.65 | 1.55 | < 0.001 |
| Financial performance | 43.39 | 2.67 | < 0.001 |
| Green revenue × Sustainability rating |  |  |  |
|    Unreported × Low | 5.34 | 1.26 | < 0.001 |
|    Unreported × High | 2.28 | 1.75 | 0.196 |
|    Much × Low | 1.82 | 1.62 | 0.264 |
|    Much × High | −1.02 | 2.14 | 0.634 |
| Green revenue × Financial performance |  |  |  |
|    Unreported × Strong | 1.05 | 1.94 | 0.590 |
|    Much × Strong | 8.28 | 2.44 | < 0.001 |
| Sustainability rating × Financial performance |  |  |  |
|    Low × Strong | 0.01 | 2.01 | 0.996 |
|    High × Strong | 3.74 | 2.48 | 0.135 |
| Green revenue × Sustainability rating × Fin. performance |  |  |  |
|    Unreported × Low × Strong | −5.03 | 2.62 | 0.058 |
|    Unreported × High × Strong | 1.67 | 2.88 | 0.564 |
|    Much × Low × Strong | −5.36 | 3.02 | 0.079 |
|    Much × High × Strong | 0.51 | 3.43 | 0.883 |
| Constant | 9.45 | 1.45 | < 0.001 |

Results of an ordinary least squares regression, with errors clustered by investor, of the investment probability on green revenue, sustainability rating, and financial performance (observations as in Table 1, $R^2 = 0.63$). Financial performance has two levels: high and low.